# Infrared study of lattice dynamics and spin-phonon and electron-phonon interactions in multiferroic TbFe$_3$(BO$_3$)$_4$ and GdFe$_3$(BO$_3$)$_4$

S.A. Klimin[1], A.B. Kuzmenko[2], M.A. Kashchenko[1,3], M.N. Popova[1]

[1] *Institute of Spectroscopy, Russian Academy of Sciences, 142190 Troitsk, Moscow, Russia*

[2] *Department of Quantum Matter Physics, University of Geneva, 1211 Geneva 4, Switzerland*

[3] *Moscow Institute of Physics and Technology (State University), 141700 Dolgoprydny, Moscow region, Russia*

**Abstract**

We present a comparative far-infrared reflection spectroscopy study of phonons, phase transitions, spin-phonon and electron-phonon interactions in isostructural multiferroic iron borates of gadolinium and terbium. The behavior of phonon modes registered in a wide temperature range is consistent with a weak first-order structural phase transition ($T_s$ = 143 for GdFe$_3$(BO$_3$)$_4$ and 200 K for TbFe$_3$(BO$_3$)$_4$) from high-symmetry high-temperature $R32$ structure into low-symmetry low-temperature $P3_121$ one. The temperature dependences of frequencies, oscillator strengths, and damping constants of some low-frequency modes reveal an appreciable lattice anharmonicity. Peculiarities in the phonon mode behavior in both compounds at the temperature of an antiferromagnetic ordering ($T_N$ = 32 K for GdFe$_3$(BO$_3$)$_4$ and 40 K for TbFe$_3$(BO$_3$)$_4$) evidence the spin-phonon interaction. In the energy range of phonons, GdFe$_3$(BO$_3$)$_4$ has no electronic levels but TbFe$_3$(BO$_3$)$_4$ possesses several ones. We observe an onset of new bands in the excitation spectrum of TbFe$_3$(BO$_3$)$_4$, due to a resonance interaction between a lattice phonon and 4$f$ electronic crystal-field excitations of Tb$^{3+}$. This interaction causes delocalization of the CF excitations, their Davydov splitting, and formation of coupled electron-phonon modes.

# I. INTRODUCTION

Multiferroic rare-earth (RE) iron borates with general formula $R$Fe$_3$(BO$_3$)$_4$ ($R$=RE or yttrium) crystallize in a noncentrosymmetric trigonal structure of the natural mineral huntite (space symmetry group $R$32). The structure contains helical chains of edge-sharing FeO$_6$ octahedra running along the $c$-axis of the crystal, interconnected by two kinds of BO$_3$ triangles and $R$O$_6$ distorted prisms [1] (Fig. 1). While in the case of Pr, Nd, and Sm iron borates the structure is described by the $R$32 ($D_3^7$) space group at all temperatures [2,3][2-5], $R$Fe$_3$(BO$_3$)$_4$ with smaller $R^{3+}$ ions undergo a structural phase transition into an also trigonal but less symmetric $P3_121(D_3^4)$ phase [6,2] at temperatures that linearly grow with diminishing the ionic radius of $R^{3+}$ [7]. Dependence of the same character holds also for the temperatures of an antiferromagnetic ordering of $R$Fe$_3$(BO$_3$)$_4$. Interestingly, the RE iron borates demonstrate a rich variety of magnetic, magnetoelectric, magnetoelastic [8-14], magnetodielectric [2,15-17], and optical [2,18-21] properties and phenomena, depending on a particular $R$ element. Some of these properties are attractive from the application point of view.

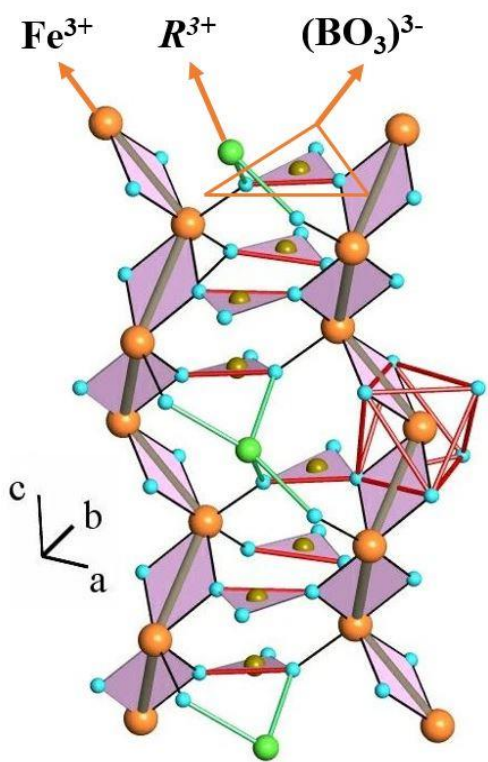

FIG. 1 (Color online). Huntite-type crystal structure of $R$Fe$_3$(BO$_3$)$_4$. Helical chains of FeO$_6$ octahedra run along the $c$-axis. Each $R$ atom interconnects three iron chains (the third not shown). Fe-O-O-Fe exchange paths between the chains are also shown.

In particular, light-induced absorption switching in a GdFe3(BO$_3$)$_4$:Nd$^{3+}$ single crystal was demonstrated, which is of interest when considering a possibility to control the optical properties of a system by light [20]. A huge magnetoelectric effect in NdFe$_3$(BO$_3$)$_4$ [10] and SmFe$_3$(BO$_3$)$_4$ [14] and a colossal magnetodielectric effect in SmFe$_3$(BO$_3$)$_4$ [17] were observed below the temperature of an antiferromagnetic ordering. Moreover, TbFe$_3$(BO$_3$)$_4$ was recently reported to exhibit a considerable quadratic magnetoelectric effect at room temperature, exceeding the one observed in the high-temperature multiferroic BiFeO3 and changing its sign upon rotation of the magnetic field by 90° [12]. These effects might be interesting for applications in spintronics and magnetic memory devices [12].

The data on TbFe$_3$(BO$_3$)$_4$ also revealed strong magnetic field effects on the dielectric constant and on the macroscopic sample length [16]. The origin of this magnetodielectric coupling was attributed to phonon mode shifts [16]. Recently, a strong interaction of a RE crystal-field excitation with a phonon and a formation of a coupled electron-phonon mode were found in the pure *R*32 compound PrFe$_3$(BO$_3$)$_4$ [21]. The electron-phonon coupling was considered as a reason for an anomalous temperature behavior of a low-frequency phonon mode in NdFe$_3$(BO$_3$)$_4$ [22]. Spin-phonon coupling effects were observed in *R*Fe$_3$(BO$_3$)$_4$ with $R$ = Eu [23], Nd, Pr, and Sm [22].

For a deeper understanding of these and other phenomena in multiferroic *R*Fe$_3$(BO$_3$)$_4$ compounds a comprehensive information on lattice modes is essential. While a thorough Raman study has been performed for the RE iron borates [2], information on infrared (IR) active phonons is scarce. Thus, Ref. [24] reports TO frequencies of all the IR active phonons in *R*Fe$_3$(BO$_3$)$_4$, ($R$ = Pr, Nd, Sm), space group *R*32, at room temperature. Ref. [25] presents reflectance spectra of *unpolarized* light from TbFe$_3$(BO$_3$)$_4$ single crystals at temperatures between 10 and 293 K, in the frequency range 20 – 600 cm$^{-1}$, which covers only a low-frequency part of the phonon spectrum. In Ref. [26], absorption spectra of HoFe$_3$(BO$_3$)$_4$ powder mixed with KBr and pressed into pellets were studied in the temperature interval from 4 to 423 K and the frequency range 30 – 1700 cm$^{-1}$. Both Ref. [25] and Ref. [26] report on changes of the phonon spectrum at the temperature $T_s$ of the structural phase transition but in none of them any changes at the temperature $T_N$ of the magnetic ordering (that could be connected with the spin-phonon coupling) were noticed.

Here we study infrared-active phonons of GdFe$_3$(BO$_3$)$_4$ and TbFe$_3$(BO$_3$)$_4$ by polarized far-infrared (FIR) reflection measurements on oriented single crystals in a wide range of temperatures (7 – 300 K) and frequencies (50 - 4400 cm$^{-1}$). Both GdFe$_3$(BO$_3$)$_4$ and TbFe$_3$(BO$_3$)$_4$ undergo a *R*32 → *P*3$_1$21 structural transformation at $T_s$ = 143 and 200 K and an antiferromagnetic ordering at $T_N$ = 32 and 40 K, respectively. While the Gd compound has no electronic crystal-field (CF) levels up to ~ 30 000 cm$^{-1}$, the Tb containing one possesses several CF levels of the ground $^7F_6$ multiplet that fall into the energy interval occupied by phonons [27].

Therefore, the considered pair of iron borates was chosen to search for spectral manifestations of the electron-phonon coupling in multiferroic RE iron borates having the $P3_121$ structure. Unlike the $R32$ phase, the $P3_121$ phase of $R$Fe$_3$(BO$_3$)$_4$ possesses three RE ions in the primitive cell, so that new effects such as Davydov splitting of 4$f$ electronic excitations could be anticipated, delivering information on RE-RE interactions and giving new features to the electron-phonon coupling.

The paper is organized in the following way. After a brief description of the experiment in Part II, in Part III we give a summary of group-theoretical results relevant to phonons and 4$f$ electronic excitations in $R$Fe$_3$(BO$_3$)$_4$ compounds. Experimental polarized FIR reflectance spectra of GdFe$_3$(BO$_3$)$_4$ and TbFe$_3$(BO$_3$)$_4$ at room temperature ($R32$ phase) and at 7 K ($P3_121$ phase) are presented in Part IV. Part V communicates on the spectra modeling and lists the phonon parameters obtained as a result of the fitting procedure. In the Part VI (Discussion), we compare our results on IR-active modes with the group-theoretical predictions and with Raman data - for $E$ modes, which are both IR and Raman active (VI A), discuss the behavior of phonon modes in the vicinity of the structural phase transition paying a special attention to quasisoft modes and effects caused by an appreciable anharmonicity (VI B), show spectral signatures of the spin-phonon interaction and of an exchange-mediated enhancement of anharmonicity (VI C), and, finally, discuss spectral manifestations of the interaction between lattice phonons and 4$f$ Tb$^{3+}$ electronic excitations and the formation of a coupled electron-phonon mode. Delocalization of the Tb$^{3+}$ crystal-field excitations and excitonic Davydov splitting are evidenced (VI D).

## II. EXPERIMENTAL DETAILS

Single crystals of gadolinium and terbium iron borates were grown on seeds from the solution-melts on the base of Bi$_2$Mo$_3$O$_{12}$, as described in Ref. [28]. Big transparent single crystals were green in color and had a good optical quality. Two polished plane-parallel plates with the $c$-axis lying in a plane were prepared. Temperature-dependent reflection spectra in the spectral range from 50 to 4400 cm$^{-1}$ and the temperature interval from 7 to 300 K were measured for the π (k ⊥ $c$, E ∥ $c$, H ⊥ $c$) and σ (k ⊥ $c$, E ⊥ $c$, H ∥ $c$) polarizations of the incident light, using a Fourier spectrometer Bruker 113 and a helium-flow cryostat. *In situ* gold evaporation was used in order to obtain the absolute reflectivity.

## III. GROUP-THEORETICAL ANALYSIS

### A. Phonons in $R$Fe$_3$(BO$_3$)$_4$ crystals

The group-theoretical analysis of the Brillouin-zone-center lattice vibrations in both the high-temperature and low-temperature phases of $R$Fe$_3$(BO$_3$)$_4$, characterized by the $R32$ ($D_3^7$) and $P3_121$ ($D_3^4$) space symmetry groups, respectively, was performed by some of us earlier [2]. This analysis included the factor-group analysis and the correlation analysis. The main results necessary for further discussion are summarized below.

The following normal optical vibrational modes exist in the two structural phases of $R$Fe$_3$(BO$_3$)$_4$:

$\Gamma_{\text{vibr}}(R32) = 7A_1(xx,yy,zz) + 12A_2(E\|z) + 19E(E\|x,E\|y; xz,yz,xy)$.

$\Gamma_{\text{vibr}}(P3_121) = 27A_1(xx,yy,zz) + 32A_2(E\|z) + 59E(E\|x,E\|y; xz,yz,xy)$.

Notations in parentheses refer to the allowed components of the electric dipole moment (IR activity) and the polarizability tensor (Raman activity). Doubly degenerate $E$ modes are polar and both IR and Raman active. The $A_2$ phonon modes are IR active in the π-polarization whereas the $E$ modes can be excited by the σ-polarized IR radiation. As the factor group is the same ($D_3$) for the two structures, the phonon symmetries (i.e., irreducible representations) remain the same, while the number of phonons increases dramatically for the low-temperature $P3_121$ phase, due to a tripling of the primitive crystal cell.

Results of the correlation analysis are summarized in Table I, where information is given on the number of phonon modes of different symmetries, depending on their origin. For instance, 12 $A_2$ modes of the $R32$ structure can be divided into 8 external and 4 internal modes (corresponding to internal vibrations of the BO$_3$ groups). External modes consist of one mode generated by the RE atom and seven modes generated by translational motions of the Fe atoms and BO$_3$ groups and by librational motions of the BO$_3$ groups. The $\nu_4$, $\nu_2$, and $\nu_3$ vibrations of a free BO$_3$ molecule generate, respectively, one, two, and one internal $A_2$ crystalline modes. Typical frequency ranges for external and internal modes are shown at the top of Fig. 2.

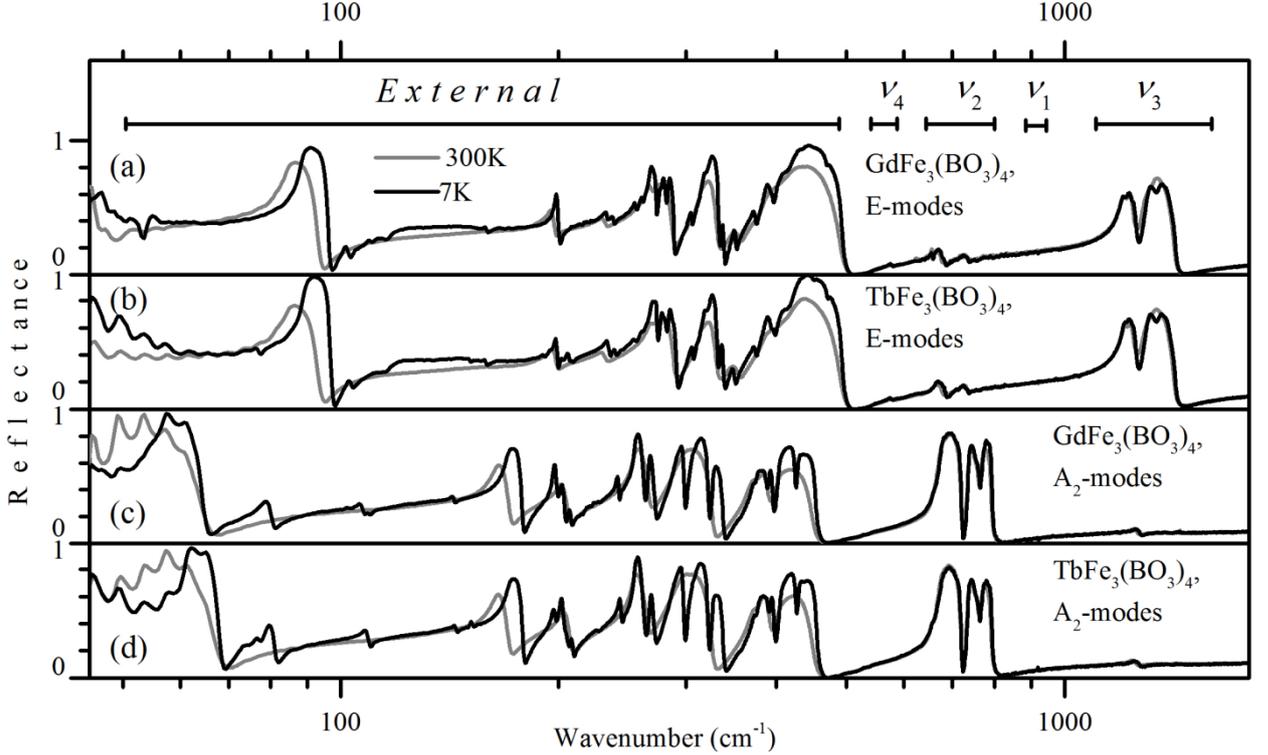

FIG. 2 (color online). Reflection spectra of (a,c) GdFe$_3$(BO$_3$)$_4$ and (b,d) TbFe$_3$(BO$_3$)$_4$ in (a,b) σ and (c,d) π polarizations at room temperature (thick gray lines) and 7 K (thin black lines). Spectral regions for external vibrations (generated by translation motions of the RE and Fe atoms and BO$_3$ groups and librational motions of the BO$_3$ groups) and internal vibrations $\nu_i$, i = 1-4, of the BO$_3$ groups are schematically shown at the top.

To study a possible interaction of phonons with crystal-field (CF) excitations associated with electronic f – f transitions within the RE subsystem, information on energies and symmetries of CF levels of $R^{3+}$ ions in $R$Fe$_3$(BO$_3$)$_4$ is essential. In the next Section, we give such information for GdFe$_3$(BO$_3$)$_4$ and TbFe$_3$(BO$_3$)$_4$.

### B. Crystal-field levels of $R^{3+}$ ions in $R$Fe$_3$(BO$_3$)$_4$

The Gd$^{3+}$ ion has the $^8S_{7/2}$ ground state with zero orbital momentum, L = 0, so that the total momentum J = S = 7/2 is of a purely spin nature. As a result, its CF splitting (which is of electrostatic nature) is zero in the first approximation for both $R$32 and $P3_121$ phases of GdFe$_3$(BO$_3$)$_4$, a very small splitting (~ 1 cm$^{-1}$) is observed due to an admixture of high-lying states with nonzero orbital momentum. The lowest excited state of Gd$^{3+}$, which is the $^6P_{7/2}$ state, is situated at ~ 32000 cm$^{-1}$, well above the frequency region occupied by phonons (~ 50 – 1000 cm$^{-1}$). Thus, no noticeable interaction of phonons with CF excitations is possible in GdFe$_3$(BO$_3$)$_4$ and we do not further analyze symmetry properties of the Gd$^{3+}$ ion in this compound.

On the contrary, the ground $^7F_6$ (L = 3, J = 6) level of $Tb^{3+}$ in $TbFe_3(BO_3)_4$ is split into CF sublevels, which occupy the energy interval ~ 500 cm$^{-1}$ [27] coincident with that for external vibrational modes (see Fig. 2). In the high-temperature $R32$ phase of $TbFe_3(BO_3)_4$, the $Tb^{3+}$ ions occupy a single $D_3$ symmetry position. The crystal field of this symmetry splits levels of free non-Kramers ions (like $Tb^{3+}$) into CF singlets and doublets characterized by the $\Gamma_1$ ($A_1$ in notations accepted for vibrations) and $\Gamma_2$ ($A_2$) non-degenerate and the $\Gamma_3$ ($E$) doubly degenerate irreducible representations (IRREPs) of the $D_3$ point symmetry group, respectively. The ground $^7F_6$ level of $Tb^{3+}$ splits into $3\Gamma_1 + 2\Gamma_2 + 4\Gamma_3$ CF levels. Optical $\Gamma_3 \rightarrow \Gamma_3$ transitions are allowed for all polarizations of light both as electric dipole (ED) and as magnetic dipole (MD); $\Gamma_1 \rightarrow \Gamma_1$ and $\Gamma_2 \rightarrow \Gamma_2$ ones are strictly forbidden; $\Gamma_1 \leftrightarrow \Gamma_2$ transitions manifest themselves in the π polarization as ED ones and in the σ polarization as MD ones; *vice versa* for the $\Gamma_1, \Gamma_2 \leftrightarrow \Gamma_3$ transitions [27].

In the low-temperature $P3_121$ phase (below $T_S$), the symmetry of the RE position lowers from $D_3$ to $C_2$ and the degeneracy of the $Tb^{3+}$ $\Gamma_3$ doublets is lifted. Wave functions of all the $Tb^{3+}$ states belong to the non-degenerate $\Gamma_1$ or $\Gamma_2$ IRREPs of the $C_2$ point symmetry group. Now, $\Gamma_1 \leftrightarrow \Gamma_2$ transitions are allowed for all polarizations both in ED and MD approximation but $\Gamma_1 \rightarrow \Gamma_1$ and $\Gamma_2 \rightarrow \Gamma_2$ transitions are allowed in the σ polarization as ED ones and in the π polarization as MD ones [27].

## IV. EXPERIMENTAL RESULTS

Reflection spectra of the two compounds, $GdFe_3(BO_3)_4$ and $TbFe_3(BO_3)_4$, for two polarizations, σ ($E\|a$) and π ($E\|c$), and at two temperatures, 300 and 7 K, are shown in Fig.2. The IR-active $A_2$ and $E$ modes are well separated from each other, in agreement with group-theoretical predictions. Understandably, spectra of the two compounds are similar (the difference in masses of Gd and Tb is about 1 % and the difference in lattice parameters for $GdFe_3(BO_3)_4$ and $TbFe_3(BO_3)_4$ is ~ 0.2% [7]). Low-temperature spectra demonstrate a great number of new phonon modes which appear due to the $R32 - P3_121$ structural phase transition [2].

Figure 3 displays reflection spectra at several temperatures together with intensity maps for both compounds in a narrow spectral range near 300 cm$^{-1}$. Temperature changes for $GdFe_3(BO_3)_4$ and $TbFe_3(BO_3)_4$ are almost identical in the spectral region presented. Both phase transitions influence the spectra. At the temperature of the structural phase transition ($T_S$ = 200 K for $TbFe_3(BO_3)_4$ and $T_S$ = 143 K for $GdFe_3(BO_3)_4$), pronounced mode shifts, splitting of some modes, and the appearance of new modes are observed. At the temperature of a magnetic ordering ($T_N$ = 40 K for $TbFe_3(BO_3)_4$ and $T_N$ = 32 K for $GdFe_3(BO_3)_4$), minor changes in the spectra arise, such as small shifts of some of the phonon frequencies.

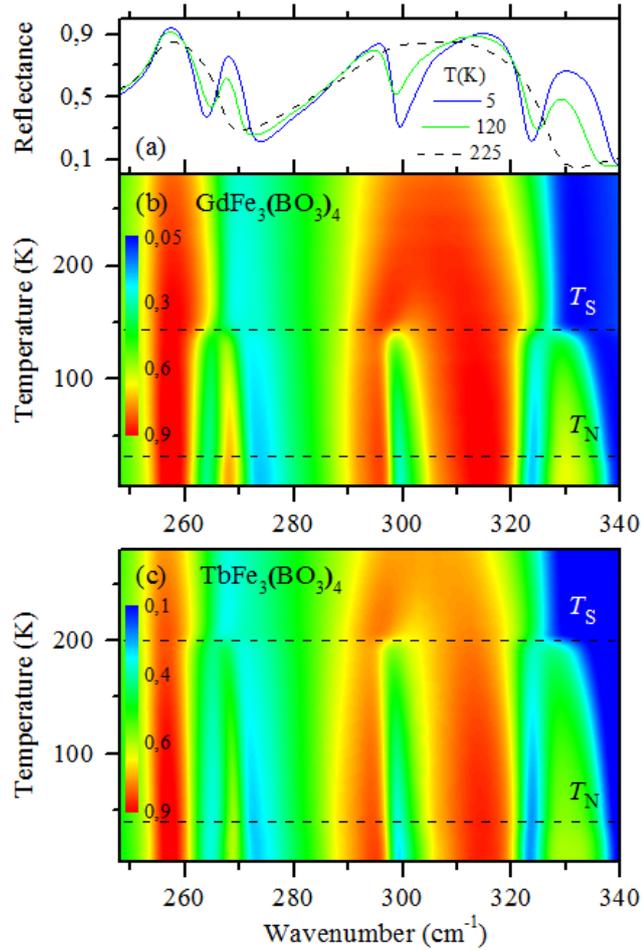

FIG. 3 (color online). The π-polarized (E∥c) (a) reflection spectra in a spectral region near 300 cm$^{-1}$ at several temperatures and (b,c) the corresponding reflection intensity maps in the frequency-temperature axes for (a,b) GdFe$_3$(BO$_3$)$_4$ and (c) TbFe$_3$(BO$_3$)$_4$.

Analogous data in the σ polarization for a spectral interval which includes the $E$ phonon mode near 200 cm$^{-1}$ are presented in Fig. 4 . In both compounds, the mode frequency demonstrates a pronounced kink at the temperature of the structural phase transition. However, at temperatures lower than ~120 K, the mode patterns are strongly different for the gadolinium and terbium compounds. In GdFe$_3$(BO$_3$)$_4$, the considered reflection peak experiences only a subtle shift and narrowing, whereas in TbFe$_3$(BO$_3$)$_4$ a new peak appears near the main one and splits into two bands below the temperature of a magnetic ordering $T_N$. These phenomena are connected to the interaction of the $E$ phonon mode near 200 cm$^{-1}$ with a CF excitation of nearly the same frequency, as will be discussed below in Section VI D.

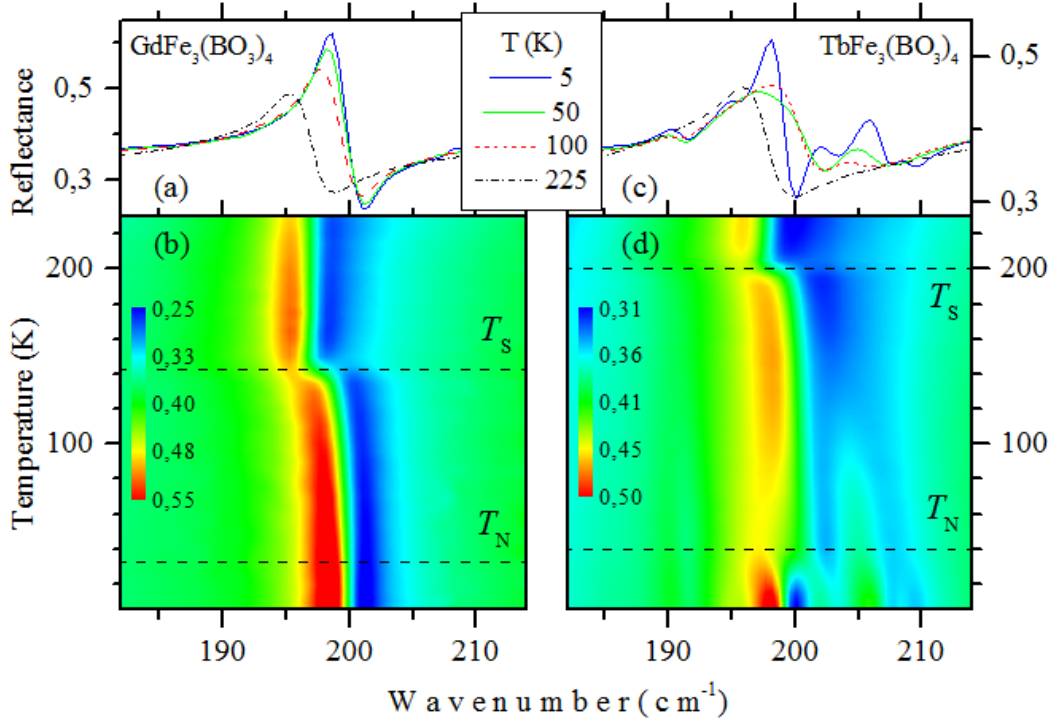

FIG. 4 (color online). The σ-polarized (E∥*a*) (a,c) reflection spectra in a spectral region of the *E*-phonon near 200 cm$^{-1}$ at several temperatures and (b,d) the corresponding reflection intensity maps in the frequency-temperature axes for (a,b) GdFe$_3$(BO$_3$)$_4$ and (c,d) TbFe$_3$(BO$_3$)$_4$.

## V. MODELING OF THE SPECTRA

Experimental spectra were fitted by the calculated ones in the framework of the Drude-Lorentz model for the dielectric function,

$$\varepsilon(\omega) = \varepsilon_\infty + \sum_i \frac{\omega_{0i}^2 \Delta\varepsilon_i}{\omega_{0i}^2 - \omega^2 - i\omega\gamma_i}, \qquad (1)$$

where $\omega_{0i}$ are the phonon transverse optical (TO) frequencies, $\gamma_i$ are the damping constants, $\Delta\varepsilon_i$ are the oscillator strengths, $\varepsilon_\infty$ is the dielectric constant at high frequencies. The reflectivity was calculated using proper Fresnel equations, taking into account a possible reflection of the infrared radiation from the back sample surface and corresponding Fabry-Perot effects. In fact, in the studied spectral range the backreflection is negligible apart from a few regions far from the optical phonons, such as 120-150 cm$^{-1}$ and below 70 cm$^{-1}$ at 7 K for E∥*a* (Figs.2a, 2b). To find experimental phonon parameters, we used a code RefFIT [29]. An example of the fitting is shown in Fig. 5 where the modeled spectrum is compared with the experimental one.

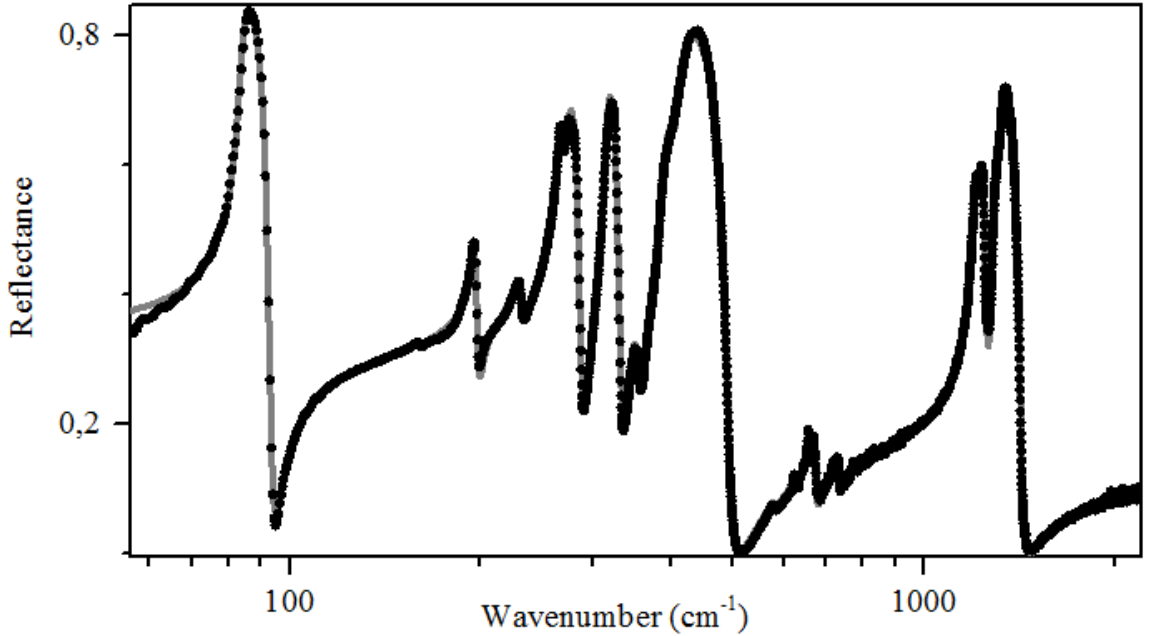

FIG. 5 (color online). Comparison of the experimental (symbols) and modeled (gray line) σ-polarized reflection spectra of GdFe$_3$(BO$_3$)$_4$ at 300 K.

Table II summarizes thus obtained $E$-phonon parameters for GdFe$_3$(BO$_3$)$_4$ and TbFe$_3$(BO$_3$)$_4$, at room temperature ($R32$ phase) and at 7 K ($P3_121$ phase). Longitudinal frequencies $\omega_{LO}$ found as maxima of the loss function Im[$-1/\varepsilon(\omega)$] [29] and available data from the Raman spectroscopy of GdFe$_3$(BO$_3$)$_4$ [2] are also given for comparison. Table III lists experimental parameters for the $A_2$ modes.

## VI. DISCUSSION

**A. IR-active phonons: Comparison with group-theoretical predictions and Raman data**

The polarized room-temperature reflection spectra of GdFe$_3$(BO$_3$)$_4$ and TbFe$_3$(BO$_3$)$_4$, that have the $R32$ structure at this temperature, are quite similar to those of $R$Fe$_3$(BO$_3$)$_4$, $R$ = Pr, Nd, Sm reported in Ref. [24]. The number of the observed modes of both $A_2$ and $E$ symmetries coincides with group-theoretical predictions, with the exception of the region of the $\nu_1$ vibration were we failed to find a corresponding $E$ phonon. We assume that, though it is not forbidden by symmetry, its intensity is very low. Indeed, the totally symmetric vibration $\nu_1$ is forbidden in the IR spectra of a free BO$_3$ molecule. It becomes allowed due to a lowering of the symmetry at the boron sites in the crystal. In such a situation, taking into account strong intramolecular bonds as compared to intermolecular, low oscillator strengths for such type of excitations are expected.

In the low-temperature $P3_121$ structure, many new IR-active phonon modes appear. However, their number is still lower than the number predicted by the group-theoretical analysis. Two mechanisms are responsible for the rise of these new modes. First, the symmetries of local positions for some of the atoms lower. In this case, the intensity of a new formerly forbidden mode $I \sim \delta^2$, where $\delta$ is a deviation from a former symmetric position. Second, the primitive cell now contains not one but three formula units, leading to an additional Davydov (factor-group) splitting proportional to the strength of the interaction between equivalent atoms inside a new primitive cell. Both these effects are, as a rule, small and that is why a part of new modes is not observed.

An interesting feature occurs when the frequency of a new weak mode falls into the TO – LO frequency interval of a strong mode of the same symmetry. In this case, the weak mode manifests itself as a dip on the top of the reflection band corresponding to the strong mode, with its longitudinal vibration at a slightly lower frequency than the nearby transverse. Such phenomenon of *inverted* phonons was first noticed in quartz [30, 31] and discussed in detail by Gervais [32]. An example of this kind can be seen in Fig. 3 where a dip appears and deepens quickly with lowering the temperature below $T_S$, on the top of a strong and broad reflection band corresponding to the $A_2$ phonon with $\omega_{TO} = 290$ cm$^{-1}$. Such behavior simulates a splitting of a strong $A_2$ phonon mode into two modes of equal intensities. One of them was identified in Ref. [25] as a new mode of the $P3_121$ phase, which, however, is not physically grounded [32].

### B. Structural phase transition

Structural phase transition in GdFe$_3$(BO$_3$)$_4$ and TbFe$_3$(BO$_3$)$_4$ was studied earlier by specific heat and Raman scattering measurements [19,2] and was shown to be a so-called weak first-order transition. A strong narrow peak in the temperature dependence of specific heat [19] and an abrupt appearance of new Raman modes at $T_S$ exhibiting a hysteretic behavior of intensities [19,2] indicated the first-order character of the phase transition. However, a strong hardening of the lowest-frequency and the most intense new Raman mode upon lowering the temperature below $T_S$ was typical for soft modes that announce a second-order structural phase transition [19,2].

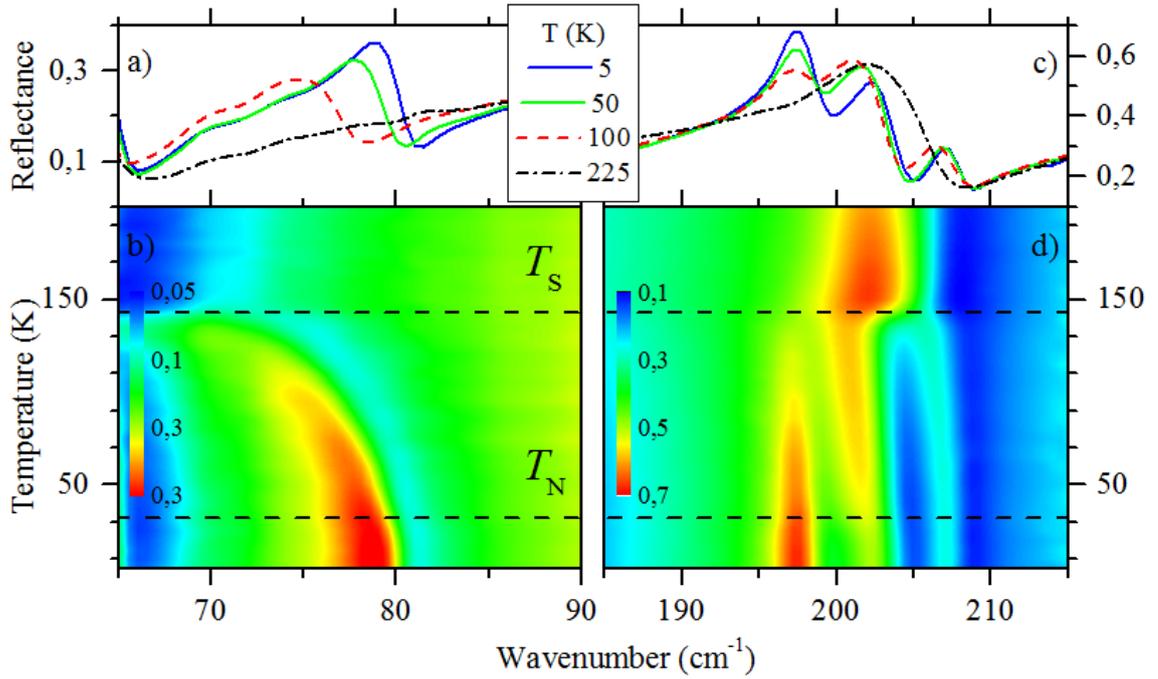

FIG. 6 (color online). (a,c) The π-polarized (E∥c) reflection spectra of GdFe$_3$(BO$_3$)$_4$ in two spectral regions at several temperatures (indicated at the top) and (b,d) the corresponding reflection intensity maps in the frequency-temperature axes.

In the IR spectra of GdFe$_3$(BO$_3$)$_4$ and TbFe$_3$(BO$_3$)$_4$, a similar soft-mode-like behavior is observed for the lowest-frequency and the strongest new mode ($A_2$ mode, Figs. 6a and 6b). It is clear that both the Raman and IR strongest new modes are associated with the biggest atomic displacements at the structural phase transition. As a detailed analysis of the structural changes shows, those occur with the BO$_3$ molecular groups [6]. In particular, BO$_3$ triangles, perpendicular to the $C_3$ axis in the $R32$ structure, tilt by ~7° in the $P3_121$ phase. Most probably, these low-frequency modes are associated just with this tilting of the BO$_3$ triangles and belong to the family of $4A_1+4A_2$ additional librational modes that appear in the low-temperature $P3_121$ structure (see Table I). Calculation of the vibrational spectrum of another iron borate, namely, of HoFe$_3$(BO$_3$)$_4$, has revealed an unstable transverse acoustical mode responsible for the $R32 \rightarrow P3_121$ structural phase transition, the frequency of which tends to zero at the Λ point of the Brillouin zone at a small change of the $z$ coordinate of the O3 oxygen atom [33]. Additional theoretical and/or experimental (especially, neutron scattering measurements) work is necessary to elucidate the nature of the just discussed $A_1$ and $A_2$ modes of GdFe$_3$(BO$_3$)$_4$ and TbFe$_3$(BO$_3$)$_4$.

Frequency *vs* temperature dependences for IR-active modes of the parent high-temperature phase demonstrate kinks at $T_S$ followed by a hardening or softening at further cooling the crystals. Typical examples can be seen in the upper panels of Figs. 7a and 7b. New phonon modes that appear at $T_S$ demonstrate similar frequency shifts, while their intensities grow (see, e.g., Figs. 3, 6c, and 6d).

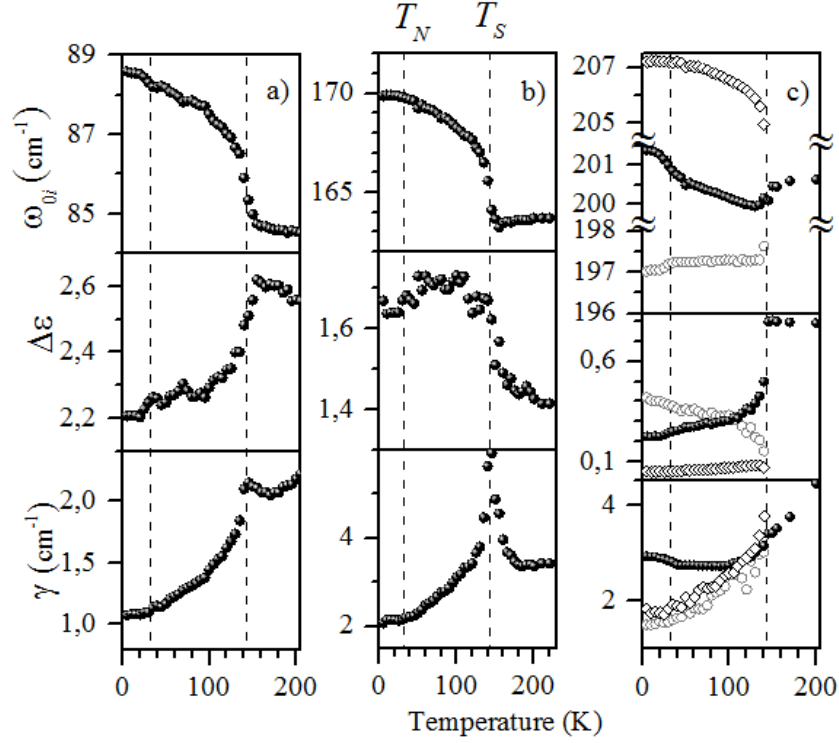

FIG. 7 (color online). Experimental data for GdFe$_3$(BO$_3$)$_4$. The temperature dependences of the frequencies $\omega_0(T)$, oscillator strengths $\Delta\varepsilon(T)$, and damping constants $\gamma(T)$ of (a) the lowest-frequency $E$ phonon mode, (b) the $A_2$ phonon mode 163.7 cm$^{-1}$ (300 K), and (c) the $A_2$ mode 200.5 cm$^{-1}$ (300 K) with two new modes in its vicinity (open circles and rhombs).

To characterize the mode frequency shifts $\Delta\omega_{0i}$ in the quasiharmonic approximation, due to the change of volume $\Delta V$ of the crystal, the mode Grüneisen parameters $g_i$ entering the relation

$$\frac{\Delta\omega_{0i}}{\omega_{0i}} = -g_i \frac{\Delta V}{V} \qquad (2)$$

are used (see, e.g., [31]). A comparison of the frequency *vs* temperature dependences of the infrared vibrational modes of quartz with thermal lattice expansion in a wide range of temperatures in the vicinity of the α → β phase transition has shown that the observed frequency shifts could be understood in term of pure volume effect with $|g_i| = 0.15 - 1.2$ in Eq. 2 [31]. For the $R$Fe$_3$(BO$_3$)$_4$ compounds, detailed data on thermal lattice expansion are absent. The only direct measurements of lattice constants were performed at room temperature and at 90 K < $T_S$ for GdFe$_3$(BO$_3$)$_4$, they yielded $V$(293 K) = 592.15 Å$^3$, $V$(90 K) = 593.73 Å$^3$, and, hence, $\Delta V/V = -2.66 \times 10^{-3}$ [6] (pay attention to the negative thermal expansion (NTE) of GdFe$_3$(BO$_3$)$_4$). Using this value and the experimentally measured frequency changes between

293 and 90 K, and assuming temperature-independent mode Grüneisen parameters (like in quartz [31]), we were able to obtain a rough estimate, according to Eq. 2, of the mode Grüneisen parameters for several isolated modes that already existed in the high-temperature parent phase of $GdFe_3(BO_3)_4$ (see Table IV). These parameters are one or two orders of magnitude greater than those for quartz [31] or silicon clathrates, $|g_i| \sim 0.13 - 1.46$ [34], We also note that all $g_i$ for the low-frequency modes of $GdFe_3(BO_3)_4$ are negative. In the quasiharmonic approximation, NTE arises from vibrational modes with negative mode Grüneisen parameters, so these modes might be responsible for NTE of $GdFe_3(BO_3)_4$ [35].

The huge values of the Gruneisen parameters, not seen in other compounds, perhaps tell in favor of their influence by the phase transition. Another reason for such unusually large $|g_i|$ could be a strong deviation from the harmonic approximation. A critical increase of modes' damping $\gamma(T)$ is observed at the approach of the transition (lower panels of Figs. 7a and 7b). In Ref. [36], such anomalous increase has been explained by anharmonic coupling of a given hard mode with a soft phonon, the linewidth of which diverges at the approach of the structural phase transition. Anharmonic couplings between phonon modes create a complex self-energy shift, the real part of which is the observed frequency shift but the imaginary part is related to the phonon damping. Appreciable anharmonicity is evident from the spectra of $GdFe_3(BO_3)_4$ presented in Figs. 6c and 6d. It manifests itself in a mutual repulsion of frequencies of the mode 200 cm$^{-1}$ and the new mode 197 cm$^{-1}$ and in a noticeable swap of intensities between these two modes. Fig.7 c shows respective $\omega_{0i}(T)$, $\Delta\varepsilon_i(T)$, and $\gamma_i(T)$ plots. Analogous data have been obtained for $TbFe_3(BO_3)_4$.

To conclude this Section, we briefly discuss a discrepancy between the values of $T_S$ reported in different publications. The first study of phase transitions in RE iron borates was performed on powder samples prepared by solid-state synthesis [7]. The temperatures $T_S = 174$ K and $T_S = 241$ K were reported for $GdFe_3(BO_3)_4$ and $TbFe_3(BO_3)_4$, respectively. Subsequent Raman scattering studies on single crystals grown by melt-solution technique using the $Bi_2Mo_3O_{12}$ based flux revealed somewhat lower temperatures $T_S = 155.7$ K and $T_S = 198.4$ K for $GdFe_3(BO_3)_4$ and $TbFe_3(BO_3)_4$, respectively [2]. This study on single crystals grown by the same method but at another growth session reports $T_S = 143$ K and $T_S = 200$ K for $GdFe_3(BO_3)_4$ and $TbFe_3(BO_3)_4$, respectively. It is known that $Bi^{3+}$ ions from the flux enter the crystal during the growth process, substituting the rare-earth ions [37]. As the ionic radius $r$ of $Bi^{3+}$ (1.03 Å) exceeds that of $Gd^{3+}$ (0.938 Å) and $Tb^{3+}$ (0.923 Å), this substitution results in a growth of the lattice constant, which in its turn diminishes $T_S$ [7]. We assume, this is the reason of different $T_S$ temperatures in differently prepared iron borate samples. Using the $T_S(r)$ dependence of Ref. [7] and the data on ionic radii, we obtain the value 4±2% for the concentration of Bi in our samples.

In the following two Sections, we concentrate on spectral signatures of interactions between lattice vibrations and the spin and electronic systems in GdFe$_3$(BO$_3$)$_4$ and TbFe$_3$(BO$_3$)$_4$

### C. Magnetic ordering and spin-lattice interactions

At the temperature of magnetic ordering $T_N$, the majority of IR-active phonon modes' frequencies exhibit weak kinks, similar to those observed and discussed for the Raman-active modes of GdFe$_3$(BO$_3$)$_4$ [2], TbFe$_3$(BO$_3$)$_4$ [2] and NdFe$_3$(BO$_3$)$_4$ [22] and IR- active modes of PrFe$_3$(BO$_3$)$_4$ [22] and EuFe$_3$(BO$_3$)$_4$ [23]. Peculiarities in the mode behavior at $T_N$ are clearly visible in Fig. 8. The "old" $A_2$ modes with room-temperature $\omega_{TO}$ = 378.1 (379.4) and 400 (403) cm$^{-1}$ for the Gd (Tb) compound demonstrate pronounced features at $T_N$ in their frequency-temperature dependences. A new mode that appears between them at the temperature of the structural phase transition $T_S$, markedly intensifies below $T_N$, at the expense of these modes and pushes them apart. The $A_2$ mode 200 cm$^{-1}$ and the new mode 197 cm$^{-1}$ (143 K) in GdFe$_3$(BO$_3$)$_4$ demonstrate a similar behavior (see the spectra of Figs. 7a and 7b and $\omega_{0i}(T)$ and $\Delta\varepsilon_i(T)$ plots of Fig.6 c.

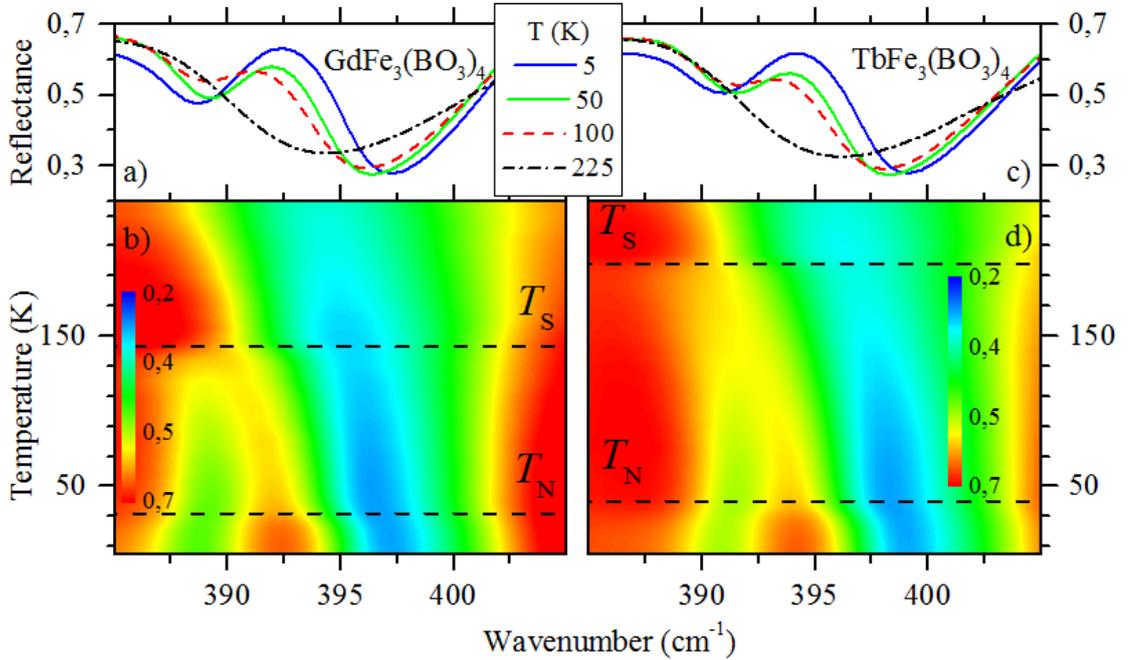

FIG. 8 (color online). The π-polarized (E||c) (a,c) reflection spectra in a spectral region 385 - 405 cm$^{-1}$ at several temperatures and (b,d) the corresponding reflection intensity maps in the frequency-temperature axes for (a,b) GdFe$_3$(BO$_3$)$_4$ and (c,d) TbFe$_3$(BO$_3$)$_4$.

The static magnetoelastic coupling causes displacements of equilibrium positions of atoms in a magnetically ordered state of iron borates, as has been proved experimentally for EuFe$_3$(BO$_3$)$_4$ [23]. This, evidently, influences normal mode frequencies. Another mechanism that could be responsible for normal mode frequency changes at the magnetic ordering is due to the phonon-induced modulation of the superexchange energies $J_{ij}(Q)$ between the spins $S_i$ and $S_j$, which, in its turn, affects the elastic constants and, hence, the phonon frequencies [38]. Such enhanced coupling between modes below $T_N$ can be also caused by an enhancement of anharmonicity due to a magnetic ordering. Indeed, the term $\Sigma <J_{ij}(Q)S_iS_j>$ in the Hamiltonian differs from zero only in a magnetically ordered state. Expansion of $J_{ij}(Q)$ in series of normal coordinates results in additional anharmonic terms.

### D. Coupled electron-phonon modes

The electron-phonon interaction in RE containing compounds originates from a modulation of the crystal field for $4f$ electrons by lattice vibrations [39]. In the energy range of phonons, $GdFe_3(BO_3)_4$ has no electronic crystal-field states of the $Gd^{3+}$ $4f$ electrons. On the contrary, the ground $^7F_6$ multiplet of $Tb^{3+}$ delivers nine CF levels in the high-temperature $R32$ phase, distributed over the energy interval 0 – 400 cm$^{-1}$ [27]. The ground state is the $\Gamma_1+\Gamma_2$ quasidoublet, the next CF levels are $\Gamma_3$ doublets at about 200 and 250 cm$^{-1}$ and a $\Gamma_1$ level 225 cm$^{-1}$ between them [27]. The most pronounced manifestations of the interaction between $4f$ electrons of $Tb^{3+}$ and lattice phonons of $TbFe_3(BO_3)_4$ are observed in the reflection spectral region near 200 cm$^{-1}$ where a $E$ phonon mode and a $\Gamma_3$ crystal-field doublet of the $R32$ phase are in resonance (see Fig. 4). At the temperature $T_S$ of the structural phase transition, the $E$ phonon experiences an abrupt frequency change but preserves the $E$ symmetry. The $\Gamma_3$ electronic excitation at the RE site of the $D_3$ symmetry (not observable in reflectance) splits into two $\Gamma_1+\Gamma_2$ singlets of the $C_2$ point symmetry group. To analyze their interaction with lattice phonons, one should establish correlations between IRREPs of the $D_3$ crystal point group and the $C_2$ local point group of the $Tb^{3+}$ ion. They take into account the presence of three $Tb^{3+}$ ions in the primitive cell of the $P3_121$ crystalline phase resulting in $\Gamma_1(C_2) \rightarrow \Gamma_1(D_3) + \Gamma_3(D_3)$; $\Gamma_2(C_2) \rightarrow \Gamma_2(D_3) + \Gamma_3(D_3)$. Thus, the local $4f$ electronic excitation transforms into the Davydov (factor-group) doublet of crystalline $4f$ excitons. One of the doublet components has the $\Gamma_3$ symmetry and can interact with $E$ phonons of the lattice, so that mixed electron-phonon excitations are formed instead of pure phonons and pure $4f$ electronic excitations.

As a result of this interaction, which grows as the thermal population of excited CF levels involved into the process diminishes, quasielectronic (i.e., involving predominantly electronic degrees of freedom) modes borrow the intensity from the quasiphonon (i.e., predominantly phonon) mode and become visible in reflectance below ~ 100 K at 191 and 205 cm$^{-1}$, at both sides of the phonon peak 198 cm$^{-1}$. It should be noted, that pure electronic f-f transitions cannot be observed in reflection, because of their very small oscillator strengths [40]. A mutual repulsion of the quasielectronic mode 205 cm$^{-1}$ and the quasiphonon mode 198 cm$^{-1}$ with decreasing the temperature is observed (compare the spectra taken at 100 and 50 K in Fig. 4), which is one more signature of the formation of coupled electron-phonon modes [41,21].

At the temperature $T_N$ = 40 K, iron magnetic moments of TbFe$_3$(BO$_3$)$_4$ order into the easy-axis antiferromagnetic structure, the direction of magnetic moments being imposed by a strong easy-axis single-ion anisotropy of the Tb$^{3+}$ ions in the crystal field of the terbium iron borate [27]. An effective magnetic field created by the Fe magnetic moments ordered along the $c$ axis, in its turn, acts upon the terbium subsystem, so that 4$f$ excitons experience shifts and splittings. In particular, the quasielectronic $\Gamma_3$ modes 191 and 205 cm$^{-1}$ and the quasiphonon mode 197 cm$^{-1}$ (coupled to electronic excitations) split into two components each. Magnetic splitting of isolated quasiphonon excitations in RE compounds has been observed and explained before [42-45]. In our case, a very complicated pattern develops, due to a mutual influence of the quasiphonon and two quasielectronic excitations. While a complete theoretical analysis would be too complicated for the present paper, we can draw some qualitative physical conclusions.

First, the observed interaction of the doubly degenerate $E$ phonon mode with 4f electronic excitations related to nondegenerate CF states of Tb$^{3+}$ in $C_2$ symmetry positions, and the splitting of these excitations at the magnetic ordering of the system, not allowed in the single-ion approximation, unambiguously point to a delocalization of the 4f electronic excitations and formation of 4f excitons [43,46].

Second, as these effects are observed only in the wave-number range of the ground $^7F_6$ multiplet of Tb$^{3+}$ in a close resonance with lattice phonons while other higher-lying multiplets are well described by the single-ion model [27], one may draw a conclusion that an interaction between the Tb$^{3+}$ ions in TbFe$_3$(BO$_3$)$_4$ (which leads to a delocalization of CF excitations) is mediated by phonons via the electron-phonon interaction. This means that electric multipole, magnetic dipole-dipole, and exchange Tb-Tb interactions, which are of comparable value for all multiplets, are weaker than the interaction caused by phonons in resonance with CF excitations.

## VII. SUMMARY

A thorough investigation of infrared-active lattice phonons has been performed for single crystals of multiferroic gadolinium and terbium iron borates, both in the high-temperature high-symmetry $R$32 and the low-temperature low-symmetry $P3_121$ structural phases, using Fourier-transform reflection spectroscopy and spectra modeling in the framework of the Drude-Lorentz model of damped oscillators. All symmetry-allowed infrared-active modes of the $R$32 phase were found and their parameters were determined, except the one $E$ mode related to the $\nu_1$ vibration of the $BO_3$ molecular unit, which is forbidden in the IR spectra of a free $BO_3$ molecule. The number of observed additional modes that appear in the low-temperature $P3_121$ structure is lower than the number predicted by the group –theoretical analysis, reasons for that are discussed. Examples of inverted phonons are given in the cases when the frequency of a new weak mode of the $P3_121$ phase falls into the TO – LO frequency interval of a strong mode.

Studies of the temperature dependences of mode frequencies, oscillator strengths, and damping constants reveal a soft-mode-like behavior for the lowest-frequency and the strongest new mode $A_2$ of the weak first-order structural phase transition ($T_S = 200$ and 143K for $TbFe_3(BO_3)_4$ and $GdFe_3(BO_3)_4$, respectively) and an appreciable anharmonicity leading to a critical increase of damping of some modes in the vicinity of $T_S$, a mutual repulsion of mode frequencies, and a noticeable swap of intensities between neighboring modes.

We have recorded spectral signatures of the interactions between the lattice, magnetic, and electronic degrees of freedom in the multiferroic title compounds. In particular, peculiarities in the phonon mode behavior at the temperature of an antiferromagnetic ordering were observed and explained by the spin-phonon interaction and an enhancement of anharmonicity due to magnetic ordering. A resonance interaction between the lattice phonon of the $E$ symmetry and $4f$ electronic crystal-field excitations of $Tb^{3+}$ in $TbFe_3(BO_3)_4$ results in delocalization of the CF excitations and their Davydov splitting and formation of coupled electron-phonon modes, which manifest themselves in the IR reflection spectra.

## ACKNOWLEDGMENTS

This work was supported by the Russian Science Foundation under Grant No 14-12-01033.

TABLE I. Number of phonon modes of different symmetries in the two phases of $R$Fe$_3$(BO$_3$)$_4$, generated by motions of specific atoms or groups of atoms, according to the group-theoretical analysis.

| Symmetry Mode | Generator Structure | Number of modes | | | | | | | |
|---|---|---|---|---|---|---|---|---|---|
| | | External | | | | BO$_3$ internal | | | |
| | | RE | Fe | BO$_3$(tr) | BO$_3$(libr) | $\nu_4$ | $\nu_2$ | $\nu_1$ | $\nu_3$ |
| $A_1$ | $R$32 | 0 | 1 | 1 | 1 | 1 | 0 | 2 | 1 |
| | $P3_121$ | 1 | 4 | 5 | 5 | 4 | 1 | 3 | 4 |
| $A_2$ | $R$32 | 1 | 2 | 3 | 2 | 1 | 2 | 0 | 1 |
| | $P3_121$ | 2 | 5 | 7 | 6 | 4 | 3 | 1 | 4 |
| $E$ | $R$32 | 1 | 3 | 4 | 3 | 3 | 1 | 1 | 3 |
| | $P3_121$ | 3 | 9 | 12 | 11 | 8 | 4 | 4 | 8 |

TABLE II. Phonon parameters for the $E$ modes in GdFe$_3$(BO$_3$)$_4$ and TbFe$_3$(BO$_3$)$_4$ at two temperatures, 300 and 7 K, obtained in this work and compared with the Raman data [2].

| | GdFe(BO$_3$)$_4$ | | | | | | | | TbFe(BO$_3$)$_4$ | | | |
|---|---|---|---|---|---|---|---|---|---|---|---|---|
| | $R$32, 300K | | | | | $P3_121$, 7K | | | $R$32, 300K | | $P3_121$, 7K | |
| | $\omega_{TO}$ | $\omega_{TO}$ [2] | $\omega_{LO}$ | $\omega_{LO}$ [2] | $\gamma$ | $\omega_{TO}$ | $\omega_{LO}$ | $\omega_{LO}$ [2] | $\omega_{TO}$ | $\omega_{LO}$ | $\omega_{TO}$ | $\omega_{LO}$ |
| External modes | | | | | | | | | | | | |
| | 83.8 | 84 | 94.1 | 95 | 2.2 | 88.5 | 96.2 | 95.5 | 83.9 | 94.1 | 88.9 | 97.8 |
| | | | | | | 101.6 | 102.3 | | | | 102.7 | 103.2 |
| | | | | | | 113 | 114 | 114 | | | 110 | 113 |
| | 159.4 | 160 | 160 | 160 | 1.9 | 158 | 158.5 | 158.4 | 159.6 | 160.2 | 158 | 158.5 |
| | 195.7 | 195 | 197.7 | 198 | 3.6 | 198.4 | 200.1 | 200 | 197 | 198.5 | 198.4 | 199.1 |
| | 230.7 | | 232 | 232 | 7.3 | 232.9 | 233.7 | 233 | 230.5 | 232 | 233.8 | 235 |
| | | | | | | 237.6 | 238.2 | | | | 239.4 | 239.6 |
| | | | | | | 255.3 | 255.6 | 254 | | | 254.9 | 255 |
| | | | | | | 260.7 | 261.2 | | | | 261.4 | 261.6 |
| | 266.8 | 270 | 271 | 270 | 5.3 | 267.4 | 273.1 | 273 | 269 | 273 | 268.9 | 274.3 |
| | | | | | | 271* | | | | | 271.2* | 270.9 |
| | 273.9 | 273 | 288.2 | 287 | 9.3 | 275.9 | 289.2 | 288 | 275.7 | 289.6 | 277.1 | 292 |
| | | | | | | 277.8* | 277 | 276 | | | 279.5* | 282 |
| | | | | | | 283.1* | 281.8 | 281 | | | 288* | 286 |
| | | | | | | 304.5 | 305.6 | 305.5 | | | 305.6 | 306.2 |
| | 315.5 | 315 | 333 | 330 | 8 | 316.4 | 333.7 | 332 | 316.1 | 332.5 | 317.3 | 332.1 |
| | | | | | | 320.4* | 319.6 | | | | 322* | 321.1 |
| | | | | | | 336.1 | 338.7 | 338 | | | 334.9 | 338.8 |
| | 349.8 | 352 | 353 | 357 | 11.5 | 349.5 | 351.4 | 351 | 347.6 | 352 | 347.2 | 349.8 |
| | | | | | | 371.9 | 374.4 | 374.5 | | | 372.4 | 374 |
| | 390.6 | 391 | 402.5 | 392.5 | 16.6 | 386.3 | 396.5 | 395 | 390.9 | 403.2 | 386.7 | 398.8 |
| | | | | | | 393* | 391.8 | 392 | | | 395.3* | 392.8 |
| | 406.3 | | 498 | 488 | 18.7 | 401.6 | 492.8 | 490 | 409.6 | 491.5 | 404.2 | 493.1 |
| | | | | | | 413* | 410.9 | | | | 415.8* | 413.7 |
| | 445* | 443 | | 443 | | 440* | 437 | | | | 455* | 453 |
| | | | | | | 473* | 470 | 472 | | | 475* | 473 |

|  | | | | | | Internal modes | | | | | | |
|---|---|---|---|---|---|---|---|---|---|---|---|---|
| $\nu_4$ | 579.2 |  | 580 | 580 | 13 | 577 | 577.9 | 576 | 579.9 | 580.5 | 576 | 576.8 |
|  |  |  |  |  |  | 596.5 | 597 | 597 |  |  | 596.7 | 597 |
|  | 630.5 | 631 | 630.8 | 633 | 13 | 631 | 631.1 |  | 631 | 631.6 | 629 | 629.2 |
|  | 669 | 670 | 674 | 676 | 13.9 | 674.3 | 680.5 |  | 671.5 | 676.2 | 677.2 | 682.6 |
| $\nu_2$ | 732.7 |  | 734.5 | 735 | 10.3 | 732.1 | 734 | 734 | 732.3 | 734.5 | 731.6 | 734.3 |
| $\nu_1$ |  |  |  |  |  |  |  | 955 |  |  |  |  |
|  |  |  | 968 |  |  |  |  | 968 |  |  |  |  |
| $\nu_3$ | 1202 | 1198 | 1215 | 1212 | 20 | 1200.6 | 1215.3 | 1212 | 1210 | 1219 | 1207.8 | 1214.2 |
|  |  |  |  | 1229 |  |  |  | 1226 |  |  |  |  |
|  |  |  |  |  |  | 1243.7* | 1243.2 |  |  |  | 1244.7* | 1244.1 |
|  | 1223 |  | 1253 | 1250 | 25.4 | 1218.6 | 1259.4 | 1257 | 1227 | 1254 | 1222.4 | 1259 |
|  | 1287 | 1280 | 1420 | 1414 | 37.2 | 1294.2 | 1422 | 1420 | 1291 | 1421 | 1296 | 1423 |
|  |  |  |  |  |  | 1341* | 1337.2 |  |  |  | 1340.9* | 1434.6 |

Footnotes for TABLE II.

* - inverted phonons

TABLE III. Phonon parameters for the $A_2$ modes in GdFe$_3$(BO$_3$)$_4$ and TbFe$_3$(BO$_3$)$_4$ at two temperatures, 300 and 7 K, obtained in this work.

|  | GdFe(BO$_3$)$_4$ | | | | | | TbFe(BO$_3$)$_4$ | | | | | |
|---|---|---|---|---|---|---|---|---|---|---|---|---|
|  | *R*32, 300K | | | *P*3$_1$21, 7K | | | *R*32, 300 K | | | *P*3$_1$21, 7K | | |
|  | $\omega_{TO}$ | $\omega_{LO}$ | $\gamma_{TO}$ | $\omega_{TO}$ | $\omega_{LO}$ | $\gamma_{TO}$ | $\omega_{TO}$ | $\omega_{LO}$ | $\gamma_{TO}$ | $\omega_{TO}$ | $\omega_{LO}$ | $\gamma_{TO}$ |
|  | | | | | | External modes | | | | | | |
|  | 48.5 | 64.1 | 1.7 | 54.6 | 64.8 | 0.7 | 52.7 | 67.5 | 2 | 60.3 | 68.1 | 0.5 |
|  |  |  |  | 78.8 | 80.6 | 3 |  |  |  | 79 | 81.3 | 3.9 |
|  |  |  |  | 107 | 107.3 | 1.7 |  |  |  | 108.6 | 109.1 | 1.7 |
|  |  |  |  | 143.2 | 143.6 | 1.5 |  |  |  | 144.2 | 144.5 | 1.2 |
|  | 163.7 | 171.5 | 4.5 | 170.2 | 178.5 | 2.5 | 164 | 171 | 4.5 | 170.7 | 179 | 2.4 |
|  |  |  |  | 197.1 | 199 | 1.6 |  |  |  | 197.1 | 198 | 1.8 |
|  | 200.5 | 206.8 | 6.4 | 201.5 | 203.8 | 3 | 202.7 | 207.7 | 5.8 | 201.1 | 206.1 | 2.2 |
|     |       |       |      | 207.2 | 208.6  | 1.8 |       |       |      | 208.5  | 209.4  | 1.8  |
|     |       |       |      | 241.5 | 242    | 1.3 |       |       |      | 243.2  | 244.6  | 1.8  |
|     | 254.1 | 267.7 | 5    | 255.1 | 263    | 2   | 254   | 266.2 | 4.4  | 254.5  | 263    | 1.4  |
|     |       |       |      | 266.4 | 272.6  | 2.9 |       |       |      | 267.8  | 267.9  | 1.6  |
|     | 289.3 | 328.7 | 9.7  | 290.4 | 323.2  |     | 287.6 | 328   | 9.2  | 287.7  | 322.8  |      |
|     |       |       |      | 305.1*| 299    |     |       |       |      | 304.4* | 298.9  |      |
|     |       |       |      | 326.8 | 338.5  | 3.6 |       |       |      | 326.2  | 338.7  | 3.5  |
|     | 368.2 | 376   | 12.2 | 371.2 | 373.5  | 6   | 370   | 376   | 11.3 | 372.5  | 374.6  | 6    |
|     | 378.1 | 390   | 7.6  | 377.3 | 388    | 11  | 379.4 | 393   | 6.5  | 377.7  | 390.1  | 9    |
|     |       |       |      | 392   | 395    | 3   |       |       |      | 394    | 398.2  | 2.8  |
|     | 400   | 450.6 | 19.8 | 402.9 | 455    |     | 403   | 451   | 18.1 | 404.6  | 456.7  |      |
|     |       |       |      | 427.2*| 425.4  |     |       |       |      | 428*   | 426.6  |      |
| Internal modes ||||||||||||
| $\nu_4$ | 675.3 | 721.6 | 8.7 | 675.6 | 721.5 | 6.6 | 675.5 | 720 | 8.5 | 675.8 | 721.5 | 7.4 |
| $\nu_2$ | 735.5 | 760.1 | 7.1 | 735.9 | 762.6 | 6.1 | 734.9 | 758.5 | 7.1 | 735.4 | 760.9 | 6.1 |
|         | 763.9 | 793   | 10  | 765.6 | 795   | 8.4 | 762.8 | 793.8 | 12  | 764.5 | 795.5 | 10.1 |
| $\nu_3$ | 1257.2| 1261  | 24.3| 1263  | 1266.7| 24.6| 1254  | 1257.3| 24.1| 1263  | 1267.5| 24.6 |

Footnotes for TABLE III.

\* - inverted phonons

TABLE IV. Mode Grüneisen parameters for several isolated vibrational modes of $GdFe_3(BO_3)_4$. Here, $\Delta\omega_0/\omega_0 = [\omega_0(293K) - \omega_0(90K)]/\frac{1}{2}[\omega_0(293K) + \omega_0(90K)]$

| Mode symmetry | $\omega_0$ (300 K) | $\Delta\omega_0/\omega_0$ | $g$ |
|---|---|---|---|
| $A_2$ | 48.5 | 0.096 | -36 |
| E | 83.8 | 0.043 | -16 |
| $A_2$ | 163.7 | 0.029 | -11 |